\documentclass[lettersize,journal]{IEEEtran}
\usepackage{amsmath,amsfonts}
\usepackage{algorithm}
\usepackage{array}
\usepackage[caption=false,font=normalsize,labelfont=sf,textfont=sf]{subfig}
\usepackage{textcomp}
\usepackage{stfloats}
\usepackage{url}
\usepackage{verbatim}
\usepackage{graphicx}
\usepackage{cite}
\usepackage{xcolor}
\usepackage{algpseudocode}
\usepackage{multirow} 

\usepackage{listings}
\usepackage{booktabs}
\usepackage{threeparttable}

\begin{document}

\title{ETHead: Generating Expressive 3D Facial Animation and Head Movement from Speech}

\author{Jiu-Cheng Xie, Jiwang Zheng, Yongkang Xia, Jian Xiong, Chi-Man Pun, Hao Gao, Feng Xu
\thanks{Jiu-Cheng Xie, Jiwang Zheng, Yongkang Xia and Hao Gao are with the School of Automation, Nanjing University of Posts and Telecommunications, Nanjing 210023, China (e-mail: jiuchengxie@gmail.com, 1024051522@njupt.edu.cn, 1225055804@njupt.edu.cn, tsgaohao@gmail.com).}
\thanks{Jian Xiong is with the School of Communications and Information Engineering, Nanjing University of Posts and Telecommunications, Nanjing 210003, China (e-mail: jxiong@njupt.edu.cn).}
\thanks{Chi-Man Pun is with the Department of Computer and Information Science, University of Macau, Taipa, Macau (e-mail: cmpun@um.edu.mo).}
\thanks{Feng Xu is with the School of Software and BNRist, Tsinghua University, Beijing 100084, China (e-mail: xufeng2003@gmail.com).}
\thanks{This work has been submitted to the IEEE for possible publication.
Copyright may be transferred without notice, after which this version may no longer be accessible. Hao Gao and Feng Xu are the corresponding authors.}
}

\markboth{Journal of \LaTeX\ Class Files,~Vol.~14, No.~8, August~2021}%
{Shell \MakeLowercase{\textit{et al.}}: A Sample Article Using IEEEtran.cls for IEEE Journals}

\maketitle



\begin{abstract}
Generating expressive 3D talking heads solely from speech remains a significant challenge due to the scarcity of high-fidelity 3D data, which limits the modeling of complex emotional motion patterns. In this paper, we introduce \textbf{E}xpressive \textbf{T}alking \textbf{Head} (ETHead), a method for generating 3D facial and head motions that vividly align with the emotional content of input speech. To overcome the data limitations, we design a self-distillation framework that leverages large-scale 2D talking videos to pre-train a specialized speech encoder. By incorporating a novel emotion-modulated probabilistic masking mechanism, this framework aligns speech representations with expressive visual dynamics, allowing the encoder to extract features highly correlated with facial and head motions directly from audio. These features are then leveraged to guide 3D generation, enriching input cues and providing explicit supervision through a joint speech-motion latent space. Extensive experiments demonstrate that ETHead substantially outperforms state-of-the-art methods. Furthermore, our motion-aligned speech encoder can serve as a transferable module, offering a general solution for enhancing expressiveness in other 3D talking head animation frameworks. The project page is available at https://verdure-oss.github.io/ETHead.github.io/.
\end{abstract}


\section{Introduction}
\IEEEPARstart{S}{peech-driven} 3D talking head is an important facial animation technique, 
enabling immersive VR/AR interaction, content creation, and digital human applications.
While speech-consistent lip motion generation is almost achieved, making the 3D head talk expressively remains an open challenge.\par

To model expressive 3D talking heads, some approaches \cite{DBLP:conf/iccv/PengWSXZH0F23,DBLP:conf/siggrapha/DanecekCTWBB23} perform speech feature disentanglement to explicitly separate emotion-related representations through cross reconstruction. However, their effectiveness is often limited by the scarcity of parallel expressive data. Alternatively, reference-based methods \cite{sun2024diffposetalk,DBLP:journals/tmm/ChenBLTWKHM25,DBLP:conf/iccv/ThambirajaHACTT23} leverage external talking clips or emotion embeddings to capture personalized speaking styles, but in practice, obtaining suitable references that cover diverse emotional expressions remains challenging. 
Overall, these works are constrained by the difficulty of acquiring specialized data that meet their respective requirements.
\par

On the other hand, 2D talking head animation has advanced significantly in recent years, enabling vivid and emotion-congruent motions. 
Beyond architectural improvements, a critical yet often overlooked factor is the data scale: the size of 2D training data has increased to tens of thousands of hours \cite{jiang2025omnihuman}. 
In contrast, training data for 3D talking heads typically ranges from several to tens of hours, largely due to the high hardware overhead of data collection. 
One possible way to bridge this gap is to apply 3D reconstruction algorithms to 2D talking videos. 
However, as noted in \cite{danvevcek2025supervising}, current techniques only ensure high precision on lab-controlled videos. 
Since high-fidelity 3D meshes are prerequisites for learning accurate speech-to-motion mappings, the scale of usable data remains severely restricted. 

\begin{figure*}[tbp]
    \centering
    \includegraphics[width=1\textwidth]{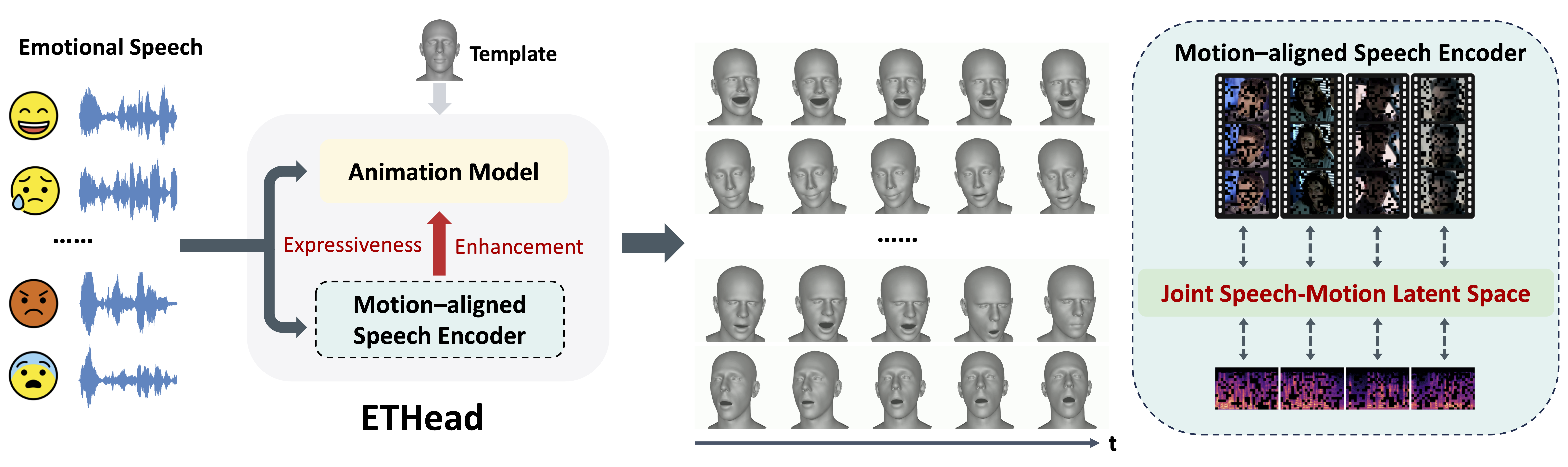}
    \caption{Given input speech, our method synthesizes emotion-coherent facial and head motions. The key is a motion-aligned speech encoder pre-trained on massive 2D talking-head data using emotion-aware masking mechanisms. This module can be seamlessly integrated into 3D talking-head frameworks to enhance their animation expressiveness.}
    \label{fig:teaser}
\end{figure*}

In this paper, we introduce ETHead (Fig.~\ref{fig:teaser}), a novel method for generating expressive 3D facial motions and head movements solely from speech. 
To address the scarcity of high-quality 3D data, ETHead leverages a hybrid training strategy that combines 3D mesh sequences with large-scale, easily accessible 2D talking videos. 
Specifically, a self-distillation learning framework is first designed to comprise dual audio and visual encoders, which are pre-trained on 2D data. 
Crucially, to capture subtle emotional dynamics, we introduce an emotion-modulated probabilistic masking mechanism that guides the encoders to focus on emotion-intensive temporal segments. 
Through such extensive multimodal pre-training, our audio encoder acquires the ability to extract rich speech representations tightly correlated with expressive movements, which is often unattainable via unimodal speech learning. 
Leveraging this powerful encoder, we facilitate 3D animation generation in two ways: by enriching the input with motion-aligned speech cues, and by providing explicit supervision via an emotion-centric, joint speech–motion latent space. 
Notably, this pre-trained speech encoder operates as an independent module, offering a transferable solution applicable to other facial animation frameworks.\par

\textcolor{black}{Experiments on standard datasets demonstrate that ETHead produces vivid facial expressions. Most notably, it significantly surpasses state-of-the-art competitors in generating natural and emotionally aligned head poses. A perceptual user study further attests to the superior realism of our results. Detailed analysis reveals that while integrating the motion-aligned speech encoder at both the input and output stages is individually effective, their synergy yields optimal improvements. Beyond our specific framework, the proposed encoder demonstrates strong transferability to other 3D head animation models, offering a versatile tool for the broader research community.}
\par

The contributions of this work are summarized as follows: \romannumeral1) We introduce ETHead, a novel method for generating expressive 3D facial motions and head dynamics consistent with the emotional content inherent in speech. \romannumeral2) We propose a motion-aligned speech encoder to capture rich visual motion priors directly from audio. Incorporating a novel emotion-modulated masking strategy, it extracts speech representations tightly correlated with expressive movements and can serve as a transferable module to empower 3D talking head animation. \romannumeral3) We conduct a comprehensive and rigorous experimental evaluation to demonstrate the effectiveness of our approach.

\section{Related Work}
\subsection{Speech-Driven 3D Facial Animation}
Speech-driven 3D talking head synthesis is a longstanding pursuit in computer graphics. Early methodologies primarily relied on rule-based or procedural animation \cite{edwards2016jali, Massaro_Cohen_Tabain_Beskow_Clark_2012, TaylorMTM12,10.1145/2522628.2522904}. While these approaches offer precise control, they necessitate extensive manual tuning and struggle to capture the nuance and variability inherent in natural speech, prompting a paradigm shift toward data-driven learning. Contemporary learning-based approaches \cite{DBLP:conf/mm/WuZ0XWW23,DBLP:conf/mm/YuPKSLSYWL24,DBLP:journals/tog/PanAS24,DBLP:journals/tvcg/SongWJLHHQ24,DBLP:journals/tvcg/ZhuangCCJLLCLL25,DBLP:journals/tvcg/SongWZLHH25,DBLP:conf/vr/LiuLZP24,DBLP:conf/cvpr/Chae-YeonHESNO25,DBLP:conf/cvpr/LiDZ0P025,DBLP:conf/mir/WangGSYLPG25,10.1145/3746027.3754933,10.1145/3746027.3755568,10.1145/3757377.3763955,DBLP:conf/wacv/NocentiniFB25,DBLP:journals/tvcg/ChaiWS025} are largely distinguished by their generative formulations. Regression-based methods \cite{DBLP:conf/icmi/HaqueY23,DBLP:conf/cvpr/CudeiroBLRB19,DBLP:conf/iccv/RichardZWTS21,DBLP:conf/eccv/FanLLXY24} are efficient but formulate synthesis as a deterministic prediction problem, which often leads to over-smoothed outputs due to regression to the mean. To capture multi-modal distributions, autoregressive approaches \cite{DBLP:conf/cvpr/FanLSWK22,DBLP:conf/siggrapha/DanecekCTWBB23,DBLP:conf/mm/PengLSXZLHF23, DBLP:journals/tmm/HanGHLLZJLZLZW25,DBLP:conf/aaai/ZhongWYW24,DBLP:conf/aaai/FuWGWCLZK24,DBLP:conf/eccv/XuGTLHLH24,DBLP:conf/eccv/NocentiniBFABD24,DBLP:conf/vr/LiuLZP24,DBLP:conf/mm/ShenXGGXD24,noh2024audio, kim2025memorytalker,yang2025stylespeaker,han2025pestalk} explicitly factorize motion over time. However, they suffer from error accumulation. Alternatively, latent variable frameworks \cite{DBLP:conf/cvpr/XingXZC0W23,DBLP:conf/iccv/ThambirajaHACTT23,DBLP:conf/cvpr/YangRCVT24,DBLP:conf/interspeech/Sung-BinCSHJNO24,DBLP:conf/mig/WuHY24,DBLP:conf/mm/WuLYD0Z24,DBLP:conf/aaai/KimCPHKKY25,DBLP:conf/aaai/LiLLM0Z25,DBLP:conf/cvpr/XieHXXXZ25,DBLP:journals/tmm/ChenBLTWKHM25} introduce stochasticity via compact latent spaces but frequently lack the representational capacity for fine-grained dynamics. Most recently, diffusion-based methods \cite{DBLP:conf/mig/StanHY23,DBLP:conf/3dim/ThambirajaPACT25,ma2024diffspeaker,DBLP:conf/icmcs/ChenWLCT25,DBLP:conf/cvpr/AnejaTDN24,DBLP:conf/siggraph/ZhaoLZQLZZYX24,DBLP:conf/ijcai/LinFWXLKPLX25,danvevcek2025supervising,DBLP:conf/cvpr/LiWCZRZZY25,10.1145/3721238.3730672,DBLP:journals/tvcg/PanLXTY25} have set a new standard by generating motion via progressive denoising. Concurrently, the scope of synthesis has expanded from strict lip synchronization to full facial animation \cite{DBLP:conf/iccv/PengWSXZH0F23,DBLP:conf/iclr/XieZLLZW25,lu2025lsf}, and more recently to the integration of head motion \cite{sun2024diffposetalk,10.1145/3721238.3730711} for enhanced emotional expressiveness. However, coordinating facial and head dynamics remains a significant challenge, largely due to the scarcity of high-quality, emotionally expressive 3D datasets. 
\textcolor{black}{To circumvent this data bottleneck, we introduce a cross-modal training strategy that leverages massive 2D talking videos to learn and transfer expressive motion priors to the 3D domain. Furthermore, rather than relying on explicit conditioning signals during generation, as seen in ProsodyTalker \cite{DBLP:conf/aaai/LiLLM0Z25} which uses fundamental frequency (F0) to drive head poses, our approach incorporates F0 solely during self-supervised pre-training to guide emotion-aware masking. By doing so, the resulting motion-aligned speech representations are strictly model-agnostic, enabling them to seamlessly enhance lip synchronization, facial expressions, and head movements across existing 3D talking head models.}

\par

\subsection{Self-Supervised Speech Representation Learning}
        Self-supervised learning (SSL) has significantly advanced speech representation learning by replacing handcrafted features with data-driven objectives. Early methods such as wav2vec 2.0 \cite{DBLP:conf/nips/BaevskiZMA20} employed contrastive learning, while HuBERT \cite{DBLP:journals/taslp/HsuBTLSM21} introduced masked prediction with offline clustering. More recent approaches have shifted toward generative masked modeling (e.g., Audio-MAE \cite{DBLP:conf/nips/000100BAGMF22}) and self-distillation via teacher–student frameworks (e.g., Data2vec \cite{DBLP:conf/icml/BaevskiHXBGA22} and emotion2vec \cite{DBLP:conf/acl/MaZYLGZ024}), with some frameworks like AV-HuBERT \cite{shi2022learning} and AV2Vec \cite{zhang2023self} extending these paradigms for visual speech recognition. Although effective for general speech understanding and recognition tasks, most existing SSL speech encoders remain agnostic to animation-oriented objectives. In expressive talking-head generation, speech information is highly non-uniform: emotionally salient segments and strong prosodic variations are closely correlated with pronounced facial and head motions \cite{livingstone2016head,krahmer2007effects}. Treating such segments equivalently to neutral or silent speech during representation learning may hinder the extraction of motion-relevant acoustic cues. \textcolor{black}{Motivated by this observation, our approach departs from conventional recognition-focused encoders by repurposing masked prediction and self-distillation paradigms to learn explicitly motion-predictive speech representations. To this end, we propose an emotion-aware masking strategy and specialized training objectives that emphasize motion-aware prosodic and emotional information, thereby prioritizing the acoustic cues essential for synthesizing expressive facial and head dynamics.}
\par

\section{Method}
Section \ref{speech encoder} first presents a speech encoder that extracts speech representations tightly coupled with expressive talking head dynamics. Next, Section \ref{talking head model} details a 3D talking head framework that synthesizes synchronized facial and head movements conditioned solely on input speech. Finally, Section \ref{emotion enhance} outlines two strategies that leverage the proposed encoder to enhance model expressiveness, encouraging the animation to align with the emotional content of the speech.
\par

\subsection{Motion-Aligned Speech Encoder}
\label{speech encoder}
\begin{figure}[tbp]
    \centering
    \includegraphics[width=\linewidth]{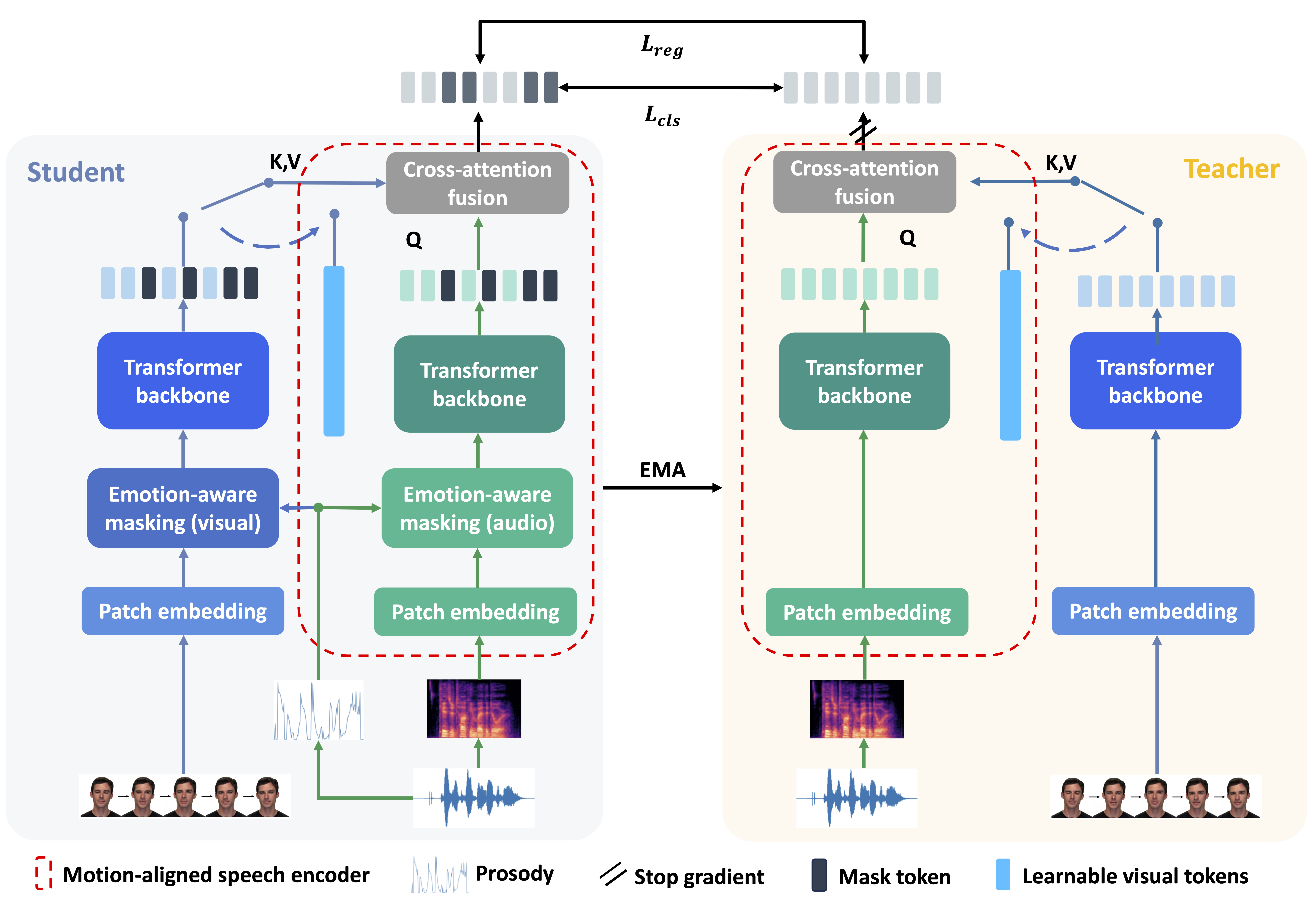} 
    \caption{\textbf{Training framework of the motion-aligned speech encoder.} Given a talking head video clip, the Student and Teacher networks separately extract modality-specific features followed by intra-network fusion. Random temporal masking is applied exclusively to the Student's features, specifically targeting segments with intense emotional dynamics identified by prosody variations. \textcolor{black}{The visual tokens are randomly dropped and replaced with learnable tokens.} The training is driven by two self-distillation objectives: $\mathcal{L}_{cls}$ aligns category predictions, while $\mathcal{L}_{reg}$ reconstructs fused tokens at masked positions.}
    \label{fig:speech-encoder}
\end{figure}

\textbf{Joint Audio-Visual Self-Distillation.}
To leverage large-scale unlabeled audio-visual 2D talking-head data, we construct a student-teacher learning framework based on the masked image modeling (MIM) and self-distillation paradigms \cite{DBLP:conf/iccv/CaronTMJMBJ21,DBLP:conf/iclr/Zhou0W0XYK22}. 
\textcolor{black}{The framework is illustrated in Fig. \ref{fig:speech-encoder} and the relevant pseudo-code is present in Algorithm \ref{alg:training3}.}

The student and teacher networks share a dual-branch architecture, with each branch processing a distinct data modality. Formally, given a paired audio-visual clip $(\boldsymbol{a},\boldsymbol{v})$, we first transform the raw audio into a Mel-spectrogram. Both the video frames and the Mel-spectrogram are then patchified and projected into sequences of patch tokens. 
\textcolor{black}{For the student network $\Theta_s$, we apply the proposed \textit{emotion-aware probabilistic masking strategies} (elaborated below) to both audio and visual data, where tokens highly correlated with emotional cues are more likely to be masked than neutral ones.} These partially masked tokens are processed by separate Transformer backbones and subsequently integrated via a cross-attention fusion module to yield the fused representation. Crucially, to enable the speech encoder to function independently during inference despite its fusion-based architecture, we introduce \textit{stochastic modality dropout}. 
During training, we randomly replace visual features with a set of learnable tokens. 
This effectively bridges the gap between training and inference, training the model to generate representations from audio signals paired with these learnable placeholders when visual data is unavailable. The teacher network $\Theta_t$ mirrors the student's architecture but processes the complete, unmasked input to provide global supervision. 

Let $\hat{\boldsymbol{z}}$ and ${\boldsymbol{z}}$ denote the fused token sequences from the student and teacher, respectively. We apply a regression loss explicitly to the masked positions:
\begin{equation}
\setlength{\abovedisplayskip}{3pt}
\mathcal{L}_{reg}=\frac{1}{M}\sum_{i=1}^{M}\|\boldsymbol{z}_i-\hat{\boldsymbol{z}}_i\|_2^2,
\setlength{\belowdisplayskip}{3pt}
\end{equation}
where $M$ represents the number of masked tokens. Additionally, we append a projection head followed by a softmax operation to both networks, producing soft probability distributions $P_s(\boldsymbol{a},\boldsymbol{v})$ and $P_t(\boldsymbol{a},\boldsymbol{v})$ from their respective outputs. A cross-entropy loss is employed to align these distributions:
\begin{equation}
\setlength{\abovedisplayskip}{3pt}
\mathcal{L}_{cls}=-\frac{1}{N}\sum_{j=1}^{N}P_t(\boldsymbol{a},\boldsymbol{v})\log P_s(\boldsymbol{a},\boldsymbol{v}),
\setlength{\belowdisplayskip}{3pt}
\end{equation}
where $N$ is the batch size. The student parameters are updated via backpropagation, while the teacher parameters are updated using an exponential moving average (EMA) of the student's parameters. 
\textcolor{black}{In addition, following DINO \cite{DBLP:conf/iccv/CaronTMJMBJ21}, we apply centering and sharpening in the teacher network to prevent representation collapse.} 
After training converges, only the speech encoder from the teacher network is retained for downstream applications.
\par

\begin{algorithm}[tbp]
{\color{black}
\caption{Training of the Motion-Aligned Speech Encoder}
\label{alg:training3}
\begin{algorithmic}
\Require Audio-visual pair $(\boldsymbol{a}, \boldsymbol{v})$, student $\Theta_s$, teacher $\Theta_t$
\Require EMA momentum $\alpha$, dropout rate $\beta$

\Repeat
    \State Compute emotional saliency $\boldsymbol{S}$ from normalized F0 trajectories
    
    \State Generate masks:
    \State \quad $\boldsymbol{M}_a \leftarrow \text{AudioMasking}(\boldsymbol{a}, \boldsymbol{S})$ 
    \State \quad $\boldsymbol{M}_v \leftarrow \text{VisualMasking}(\boldsymbol{v}, \boldsymbol{S}, \boldsymbol{a})$ 
    
    \State Forward pass:
    \State \quad $\hat{\boldsymbol{z}}, P_s \leftarrow \Theta_s(\boldsymbol{a} \odot \boldsymbol{M}_a, \boldsymbol{v} \odot \boldsymbol{M}_v, \beta)$
    \State \quad $\boldsymbol{z}, P_t \leftarrow \Theta_t(\boldsymbol{a}, \boldsymbol{v}, \beta)$
    
    \State Compute loss:
    \State \quad $\mathcal{L} = \mathcal{L}_{reg}(\hat{\boldsymbol{z}}, \boldsymbol{z}) + \mathcal{L}_{cls}(P_s, P_t)$
    
    \State Optimization:
    \State \quad $\Theta_s \leftarrow \text{Optimizer}(\nabla_{\Theta_s} \mathcal{L})$
    \State \quad $\Theta_t \leftarrow \alpha \Theta_t + (1 - \alpha) \Theta_s$

\Until{convergence of $\Theta_s$, $\Theta_t$}
\end{algorithmic}
}
\end{algorithm}
\textcolor{black}{\textbf{Design Philosophy.}
This framework establishes a critical informational asymmetry during training: for the audio modality, the teacher views the full context, while the student learns to reconstruct the comprehensive representation from partially observed inputs with attenuated emotional cues.
From a representation learning perspective, solving this masked reconstruction task gradually enables the model to focus on the emotion-rich features embedded in the observable input and to model contextual dependencies. More crucially, integrating paired visual priors offers a unique advantage.
Because emotionally expressive speech is not always accompanied by salient facial or head movements, this visual sparsity forces the alignment process to act as a ``cross-modal focuser". Specifically, forcing the student's audio-visual fused output to match the teacher's full representation compels the speech encoder to concentrate on segments where expressive speech and visual kinematics co-occur. Consequently, the model learns to extract subtle speech patterns intrinsically linked to facial and head movements.
We hypothesize, and empirically verify in Section \ref{ablation_study}, that such cross-modal dependency captures robust, motion-aligned speech features unattainable when training on audio signals in isolation.
Furthermore, by leveraging large-scale unlabeled 2D audio-visual resources, our approach holds the potential to benefit from scaling effects as data availability continues to grow.
}
\par


\textcolor{black}{\textbf{Prosody-Driven Emotion-Aware Masking.}
Since speech emotion is inherently manifested through prosody, we utilize fundamental frequency (F0), the primary acoustic measure of pitch, to track affective dynamics. By normalizing F0 trajectories against the statistics of neutral speech, we generate a shared emotion saliency profile that captures prosodic intensity fluctuations. Guided by this profile, we propose emotion-aware masking strategies to force the model to infer affective dynamics from surrounding contexts. 
Specifically, the audio branch employs a non-uniform masking strategy that biases sampling toward emotionally salient time steps. Similarly, the visual branch extends this principle by increasing the masking probability of visual patches that strongly correlate with the global prosody sequence.
Both mechanisms incorporate stochasticity to prevent overfitting and ensure training diversity. Refer to the Appendix for detailed implementations.
} 
\par

\begin{figure*}[tbp]
    \centering
    \includegraphics[width=0.90\textwidth]{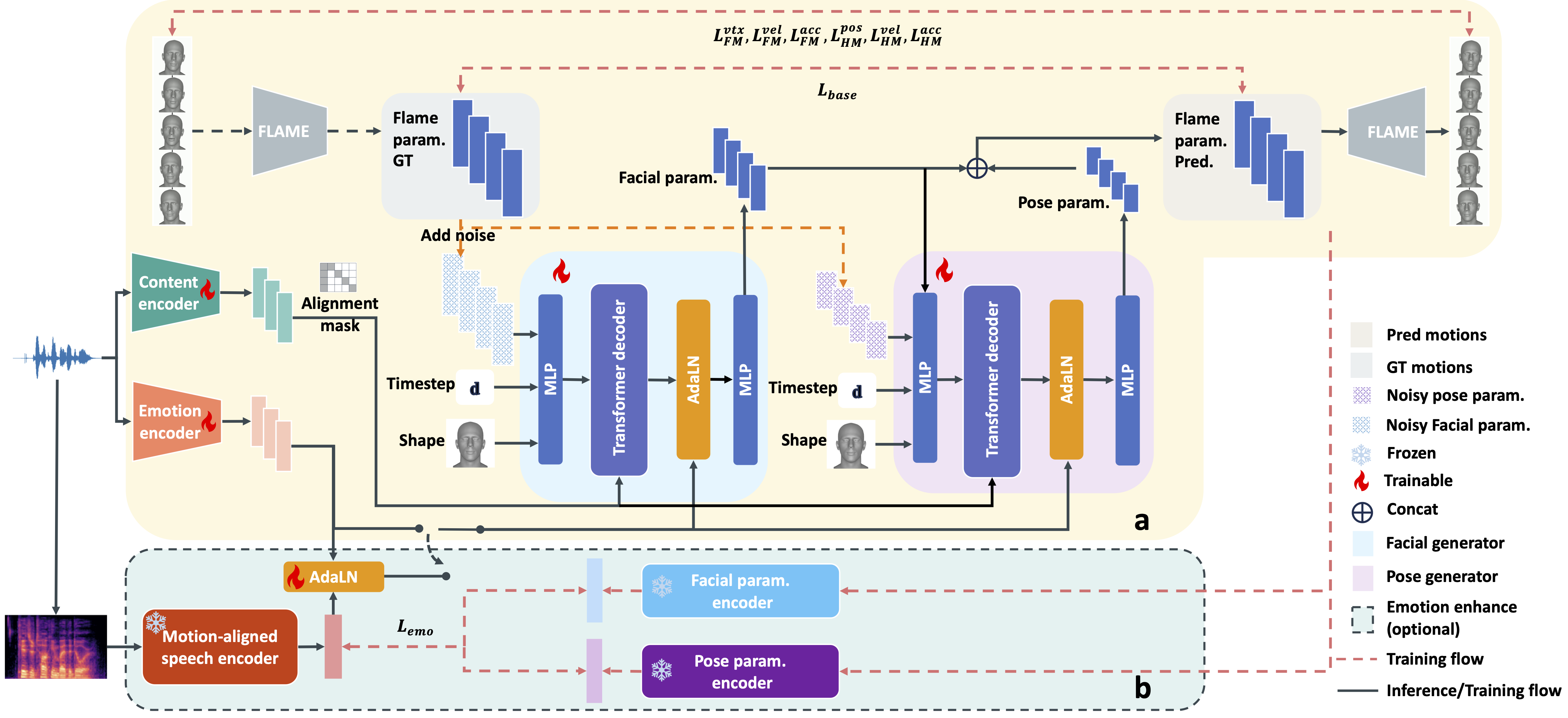} 
    \caption{\textbf{Overview of the proposed framework.} (a) Generation Pipeline: The model extracts linguistic content and emotional features from input speech to condition two diffusion-based generators. In a cascaded manner, synthesized facial motion parameters are combined with speech features to drive the subsequent head motion generator. Training is supervised by reconstruction and kinematic consistency objectives. (b) Auxiliary Module: An optional motion-aligned speech encoder enhances the baseline through input feature modulation and output-level supervision within a pre-trained speech-motion latent space.}
    \label{fig:over_pipeline}
\end{figure*}
\subsection{Speech-Driven 3D Talking Head Model}
\label{talking head model}
\textbf{Formulation.} 
Our model operates by taking speech as the driving signal, conditioned on a neutral template mesh required by the underlying FLAME topology \cite{FLAME:SiggraphAsia2017}. The network outputs sequential parameters governing both facial dynamics and head pose dynamics. Specifically, we represent facial motion using a combination of expression coefficients and jaw pitch, while head pose is defined by neck rotation. Notably, we convert the 3D axis-angle neck rotation into a 6D representation \cite{DBLP:conf/cvpr/ZhouBLYL19}. This continuity-preserving representation is empirically proven to facilitate stable network training.
\par

\textbf{Architecture.}
Fig. \ref{fig:over_pipeline}(a) depicts the framework of our method. For speech representation, we utilize WavLM \cite{DBLP:journals/jstsp/ChenWCWLCLKYXWZ22} to extract content features and Emotion2vec \cite{DBLP:conf/acl/MaZYLGZ024} to capture emotion cues. Given the distinct statistical properties of expression and pose parameters, we design independent generators for each task. Both adhere to the diffusion-based generative paradigm \cite{ho2020denoising}, learning to recover clean motion parameters from noisy latents. During training, noisy inputs are produced via a forward diffusion process following a fixed schedule, whereas during inference, inputs are sampled from a standard Gaussian distribution. Structurally, both generators share an identical architecture comprising two MLPs, a Transformer decoder, and an Adaptive Layer Normalization (AdaLN) module. The initial MLP encodes the noisy motion parameters, diffusion timestep, and template mesh. These inputs are aggregated into high-dimensional embeddings, which then interact with speech content features via cross-attention. To preserve temporal alignment, we apply an alignment mask \cite{DBLP:conf/cvpr/FanLSWK22} that restricts motion tokens to attend only to synchronous speech frames. Subsequently, the decoder output is modulated by speech emotion features via AdaLN \cite{DBLP:conf/iccv/HuangB17} and projected by a final MLP to yield the denoised motion parameters. To model the coupling between facial dynamics and head movements, we explicitly condition the head pose generator on the output of the expression generator.
\par

\textbf{Training Losses.}
To handle speech of arbitrary lengths, we adopt the sliding window strategy utilized in \cite{sun2024diffposetalk}. Specifically, we employ a window with a fixed length of $T_w$ frames. To ensure smooth transitions during the sliding process, we set a fixed overlap of $T_p$ frames between consecutive windows. Let $\boldsymbol{X}_{-T_p:T_w}$ denote the ground truth parameters for facial motion and head pose, and $\hat{\boldsymbol{X}}_{-T_p:T_w}$ denote the denoised clean version predicted by the generators. Our fundamental objective is parameter reconstruction, defined as:
\begin{equation}
\setlength{\abovedisplayskip}{3pt}
    \mathcal{L}_{base}=\lVert \boldsymbol{X}_{-T_p:T_w} - \hat{\boldsymbol{X}}_{-T_p:T_w} \rVert_2^2.
\setlength{\belowdisplayskip}{3pt}
\end{equation}
Since basic reconstruction is insufficient for effective supervision, we introduce additional constraints at the geometric level. Specifically, we convert the facial motion parameters into mesh vertices in the canonical pose (zero pose), denoted as $\boldsymbol{M}$. We then construct losses to enforce consistency in position and velocity, while imposing a smoothness constraint on acceleration, as formulated in \cite{DBLP:conf/siggrapha/DanecekCTWBB23,sun2024diffposetalk}:
\begin{equation}
\setlength{\abovedisplayskip}{3pt}
    \mathcal{L}_{FM}^{vtx}=\lVert \boldsymbol{M}_{-T_p:T_w} - \hat{\boldsymbol{M}}_{-T_p:T_w} \rVert_2^2,
\setlength{\belowdisplayskip}{3pt}
\end{equation}
\begin{equation}
\setlength{\abovedisplayskip}{3pt}
\setlength{\belowdisplayskip}{3pt}
\begin{aligned}
    \mathcal{L}_{FM}^{vel} &= \lVert ( \boldsymbol{M}_{-T_p+1:T_w}-\boldsymbol{M}_{-T_p:T_w-1}) \\
    &\quad - (\hat{\boldsymbol{M}}_{-T_p+1:T_w}-\hat{\boldsymbol{M}}_{-T_p:T_w-1})\rVert_2^2,
\end{aligned}
\end{equation}
\begin{equation}
\setlength{\abovedisplayskip}{3pt}
    \mathcal{L}_{FM}^{acc}=\lVert\hat{\boldsymbol{M}}_{-T_p+2:T_w} - 2\hat{\boldsymbol{M}}_{-T_p+1:T_w-1} + \hat{\boldsymbol{M}}_{-T_p:T_w-2}\rVert_2^2.
\setlength{\belowdisplayskip}{3pt}
\end{equation}
\par

Regarding the constraints on head pose, we first employ an objective based on the geodesic distance \cite{hempel20226d}:
\begin{equation}
    \mathcal{L}_{HM}^{pos}=\cos^{-1} \left( \frac{tr(\hat{R} R^T) - 1}{2} \right),
\end{equation}
where $R$ and $\hat{R}$ are rotation matrices of ground truth and predicted head pose, respectively. Here, $tr(\cdot)$ denotes the trace of a matrix, and $cos^{-1}(\cdot)$ represents the inverse cosine function. The velocity and acceleration constraints, $\mathcal{L}_{HM}^{vel}$ and $\mathcal{L}_{HM}^{acc}$, follow the same formulation as their counterparts for facial motion. 
\par

Consequently, the total loss $\mathcal{L}$ for model supervision is formulated as:
\begin{equation}
\setlength{\abovedisplayskip}{3pt}
\mathcal{L}_{FM}=\lambda_{FM}^{vtx}\mathcal{L}_{FM}^{vtx} + \lambda_{FM}^{vel}\mathcal{L}_{FM}^{vel} + \lambda_{FM}^{acc}\mathcal{L}_{FM}^{acc},
\setlength{\belowdisplayskip}{3pt}
\end{equation}
\begin{equation}
\setlength{\abovedisplayskip}{3pt}
\mathcal{L}_{HM}=\lambda_{HM}^{pos}\mathcal{L}_{HM}^{pos} + \lambda_{HM}^{vel}\mathcal{L}_{HM}^{vel} + \lambda_{HM}^{acc}\mathcal{L}_{HM}^{acc},
\setlength{\belowdisplayskip}{3pt}
\end{equation}
\begin{equation}
\setlength{\abovedisplayskip}{3pt}
    \mathcal{L}=\mathcal{L}_{base} + \mathcal{L}_{FM} + \mathcal{L}_{HM},
\setlength{\belowdisplayskip}{3pt}
\end{equation}
where the $\lambda$ terms serve as weighting factors to balance the objectives.
\par

\subsection{Dual-Role Expressiveness Enhancement}
\label{emotion enhance}
Leveraging our pre-trained speech encoder's ability to capture motion-critical expressive features, we integrate it into the proposed 3D talking head model at both the input and output levels. While each integration is effective individually, Section \ref{ablation_study} demonstrates that their combination yields the most significant performance improvements.
\par

\textbf{Input Enhancement.}
Our speech encoder is pre-trained on audio-visual data with emotion-aware masking, enabling it to extract speech representations intrinsically linked to expressive facial and head dynamics. In contrast, standard audio-only emotion encoders (e.g., emotion2vec \cite{DBLP:conf/acl/MaZYLGZ024}) focus solely on acoustic emotional semantics. While effective for classification, such encoders often overlook specific acoustic patterns that trigger visual expressiveness due to a lack of cross-modal grounding. Consequently, relying exclusively on audio-only features is insufficient for generating vivid motion. To bridge this gap, we utilize our visually-grounded speech encoder to enrich the input signal. Specifically, we employ features from our encoder to modulate the general emotional representations derived from the audio-only encoder via AdaLN \cite{DBLP:conf/iccv/HuangB17}. This strategy effectively injects motion-correlated cues into the semantic emotional embeddings, thereby facilitating a more accurate mapping from speech to expressive movements.
\par

\textbf{Output Supervision.} 
Drawing upon the concept of joint speech-motion representation learning introduced in \cite{DBLP:conf/aaai/KimCPHKKY25,DBLP:conf/cvpr/Chae-YeonHESNO25,DBLP:conf/iclr/XieZLLZW25}, we construct two joint latent spaces to implement output supervision. Specifically, we extract motion-critical expressive features using our speech encoder $E_{sp}$, while simultaneously employing two separate extractors $E_{fm}$ and $E_{hm}$ to encode the paired facial and head motion parameters. These two motion extractors are learned via contrastive learning: speech and kinematic features from the same time window are treated as positive pairs, whereas those from different time windows form negative pairs. The training objective \cite{oord2018representation} encourages high similarity between positive pairs while pushing apart negative ones. Through this alignment with our motion-salient acoustic representations, the motion extractors are implicitly conditioned to prioritize expressive attributes over neutral ones. Once converged, these extractors, along with the speech encoder, serve as frozen supervisors for the 3D talking head model, regularizing the generated motion to align with the expressive motion cues. The relevant loss is written as:
\begin{equation}
\setlength{\abovedisplayskip}{3pt}
    \mathcal{L}_{emo}=2-cos(E_{sp}(\boldsymbol{a}),E_{fm}(\boldsymbol{e}))-cos(E_{sp}(\boldsymbol{a}),E_{hm}(\boldsymbol{p})),
\setlength{\belowdisplayskip}{3pt}
\end{equation}
where $\boldsymbol{e}$ denotes the facial motion sequence and $\boldsymbol{p}$ represents the head pose sequence. $cos(\cdot , \cdot)$ computes the cosine similarity among two feature vectors.   
\par

\section{Experiments}
\subsection{Experimental Setup}

\textbf{Datasets.} We construct our datasets to serve two distinct phases: speech encoder pre-training and 3D supervision. For the pre-training of the motion-aligned speech encoder, we leverage two public 2D audio-visual datasets, CelebV-HQ \cite{DBLP:conf/eccv/ZhuWZJTZLL22} and CelebV-Text \cite{DBLP:conf/cvpr/YuZJL0W23}. To investigate the impact of data scale on representation learning, we curate three subsets of increasing magnitude. The \textit{Small} set comprises 6 hours sampled from CelebV-HQ, while the \textit{Medium} set utilizes the full 68-hour dataset. The \textit{Large} set further expands the scale to 250 hours by combining CelebV-HQ with portions of CelebText. Regarding 3D supervision, we address the lack of expressive 3D talking head data by generating pseudo-ground truth via the SMIRK \cite{DBLP:conf/cvpr/RetsinasFDARBM24} reconstruction framework. We primarily target the talking-head segments of RAVDESS \cite{livingstone2018ryerson} and MEAD \cite{DBLP:conf/eccv/WangWSYWQHQL20}, where controlled laboratory conditions ensure satisfactory reconstruction quality after filtering and smoothing. Furthermore, we extend this pipeline to the HDTF dataset \cite{DBLP:conf/cvpr/ZhangL0F21}, which consists of high-resolution talking records in daily-life scenarios, characterized by natural, spontaneous motion rather than emphasized expressiveness.
\par

\textbf{Evaluation Protocols.}
The in-domain evaluation is performed on 3D-RAVDESS to assess the fidelity of emotional synthesis under a ``seen-subject, unseen-utterance" condition. Since the dataset contains 24 actors delivering two fixed sentences across eight emotions, we adopt a specific partitioning strategy to separate linguistic content from identity. We construct the training set to include the complete data from 21 actors, augmented by samples of the first sentence from the remaining three actors. The validation set then consists of the samples corresponding to the second sentence from one of these three actors, while the test set comprises the second sentence samples from the other two. This protocol ensures that the model is exposed to the facial structures and expressive styles of the test subjects during training, yet it is evaluated on linguistic content it has not previously encountered for those specific identities. This allows for a direct assessment of the model's capability to reproduce actor-specific emotional nuances. The \textcolor{black}{out-of-domain} evaluation follows a typical ``unseen-subject, unseen-utterance" setting to test generalization capabilities. In this scenario, we train and validate the model on the combined 3D-RAVDESS and 3D-HDTF datasets, while testing is conducted on 3D-MEAD. As 3D-RAVDESS is limited to two fixed sentences, the inclusion of 3D-HDTF is critical to cover a broader vocabulary. This combination ensures the model is trained on sufficient linguistic variability, enabling meaningful generalization to the diverse utterances found in the 3D-MEAD dataset.
\par



\textbf{Metrics and Baselines.}
To quantitatively evaluate the quality of synthesized 3D facial motion, we adopt three standard metrics following previous conventions: Lip Vertex Error (LVE) \cite{DBLP:conf/iccv/RichardZWTS21}, Emotional Vertex Error (EVE) \cite{DBLP:conf/iccv/PengWSXZH0F23}, and Upper-Face Dynamics Deviation (FDD) \cite{DBLP:conf/cvpr/XingXZC0W23}. Among these, LVE primarily reflects lip synchronization, while EVE and FDD are indicative of emotional congruency. For head pose dynamics evaluation, we employ Beat Alignment (BA) \cite{sun2024diffposetalk} and Frechet Inception Distance (FID) \cite{DBLP:conf/cvpr/SiyaoYGLW0L022}. BA quantifies rhythmic consistency between each synthesized head-movement beat and its nearest counterpart in the ground-truth sequence. FID assesses the realism of the generated head poses. 

We compare ETHead against three state-of-the-art methods: DiffPoseTalk \cite{sun2024diffposetalk}, LSF-Animation \cite{lu2025lsf}, and DEEPTalk \cite{DBLP:conf/aaai/KimCPHKKY25}. To ensure fair comparisons under our audio-only setting, we adapt the first two methods. Originally, DiffPoseTalk requires an explicit speaking style extracted from an auxiliary talking video. We evaluate two variants of it: a default version that omits this style input for a strict, fair comparison, and an oracle version that retains it. For the oracle version, the input speech and the auxiliary video share the same identity and emotion but use different utterances to avoid trivial ground-truth leakage. This version serves as an upper bound, characterizing how well an audio-only encoder can perform compared to a method given explicit style cues. Additionally, we adapt LSF-Animation by extending its output dimensionality to include neck rotation parameters, enabling head pose animation.
We retrain DiffPoseTalk and LSF-Animation based on their official codebases, modified where necessary to incorporate our adaptations mentioned above, and evaluate them on both the 3D-RAVDESS and 3D-MEAD datasets. DEEPTalk, in contrast, is evaluated exclusively on 3D-MEAD using its official pre-trained checkpoint. Crucially, while testing on 3D-MEAD is a strict out-of-domain evaluation for ETHead and the retrained baselines, the official DEEPTalk model was trained directly on this dataset. Consequently, DEEPTalk is treated as an in-domain reference to evaluate the zero-shot generalization capabilities of our method, rather than a direct baseline.




\par

\begin{figure*}[tbp]
    \centering
    \includegraphics[width=0.75\textwidth]{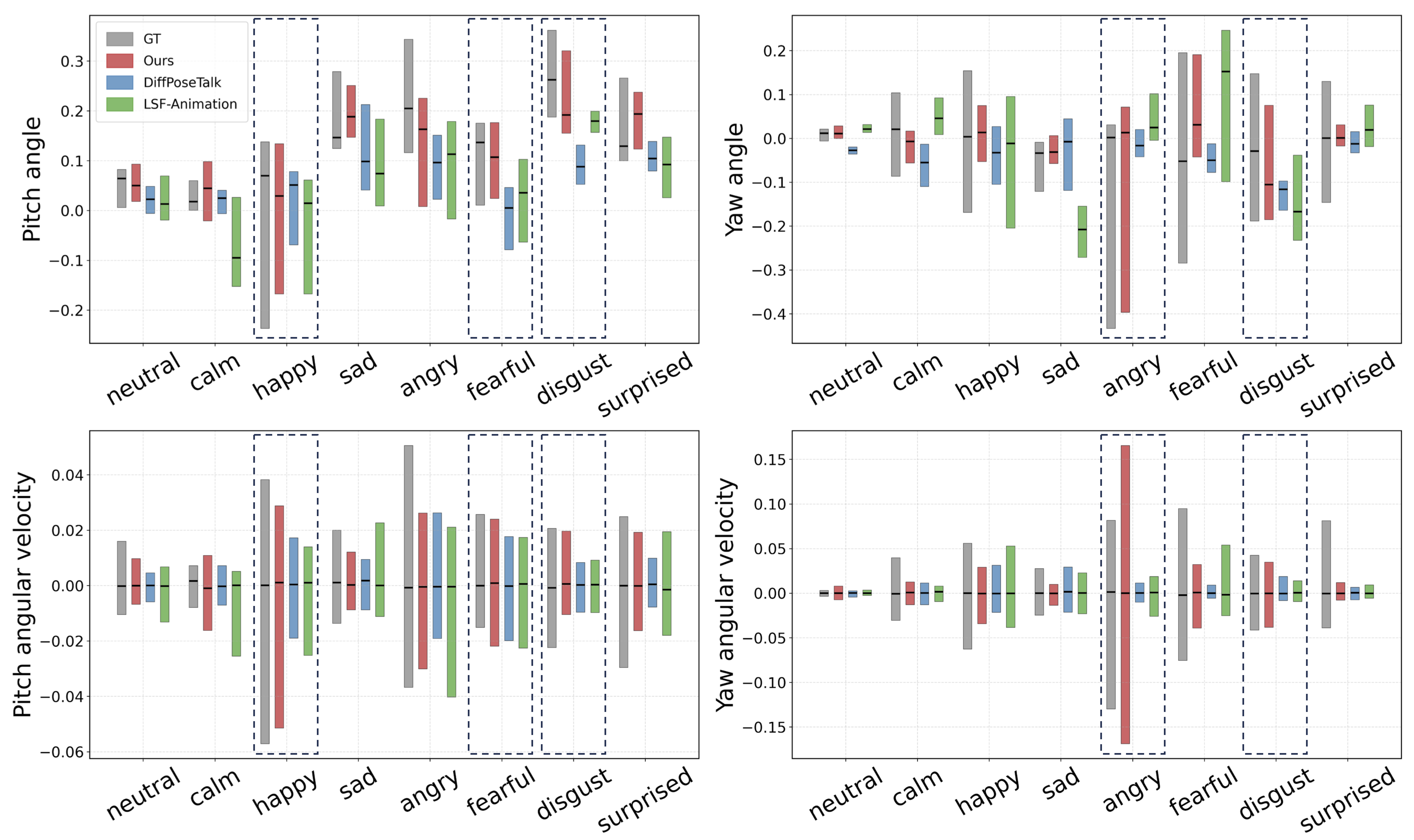}
    \caption{\textbf{Statistical analysis of head pose dynamics.} Results are shown for a specific speaker in the 3D-RAVDESS test set. The boxes indicate the full range (minimum to maximum), with the center line denoting the median. Our method produces pitch, yaw, and their angular velocity statistics that closely match the Ground Truth, significantly outperforming baselines in preserving speaker-specific styles under emotional conditions. Key comparisons are marked by dashed boxes.}
    \label{fig:statistics}
\end{figure*}
\subsection{Quantitative Comparisons}
\textbf{In-Domain Evaluation.}
Quantitative results \textcolor{black}{from a single run} under the in-domain setting are summarized in Table~\ref{tab:in domain}. Our method achieves lower errors than baseline methods on all metrics. Specifically, on the EVE, FID, and LVE metrics, it exhibits substantially larger improvements, indicating more accurate facial motion modeling, more natural head pose generation, and improved lip synchronization, respectively. 
\textcolor{black}{Notably, on most metrics, our method even outperforms the oracle version of DiffPoseTalk (marked with $\ast$), which utilizes the explicit talking style of the target speaker as an input condition. This impressive result demonstrates that our model can effectively deduce and preserve speaker-specific styles and rich emotional characteristics directly from the speech audio itself, bypassing the need for explicit style guidance.}
We further analyze the statistics of the head pose dynamics of the selected test speaker. As shown in Fig. \ref{fig:statistics}, our method produces head rotation ranges and angular velocities that are closer to the ground truth than those of the baselines across different emotions. This suggests that our approach more accurately captures speaker-specific head movement patterns under various emotional states. Overall, these results support the effectiveness of our design, which jointly models motion-aligned speech cues and 3D facial and head motion dynamics.
\begin{figure*}[tbp]
    \centering
    \includegraphics[width=0.90\textwidth]{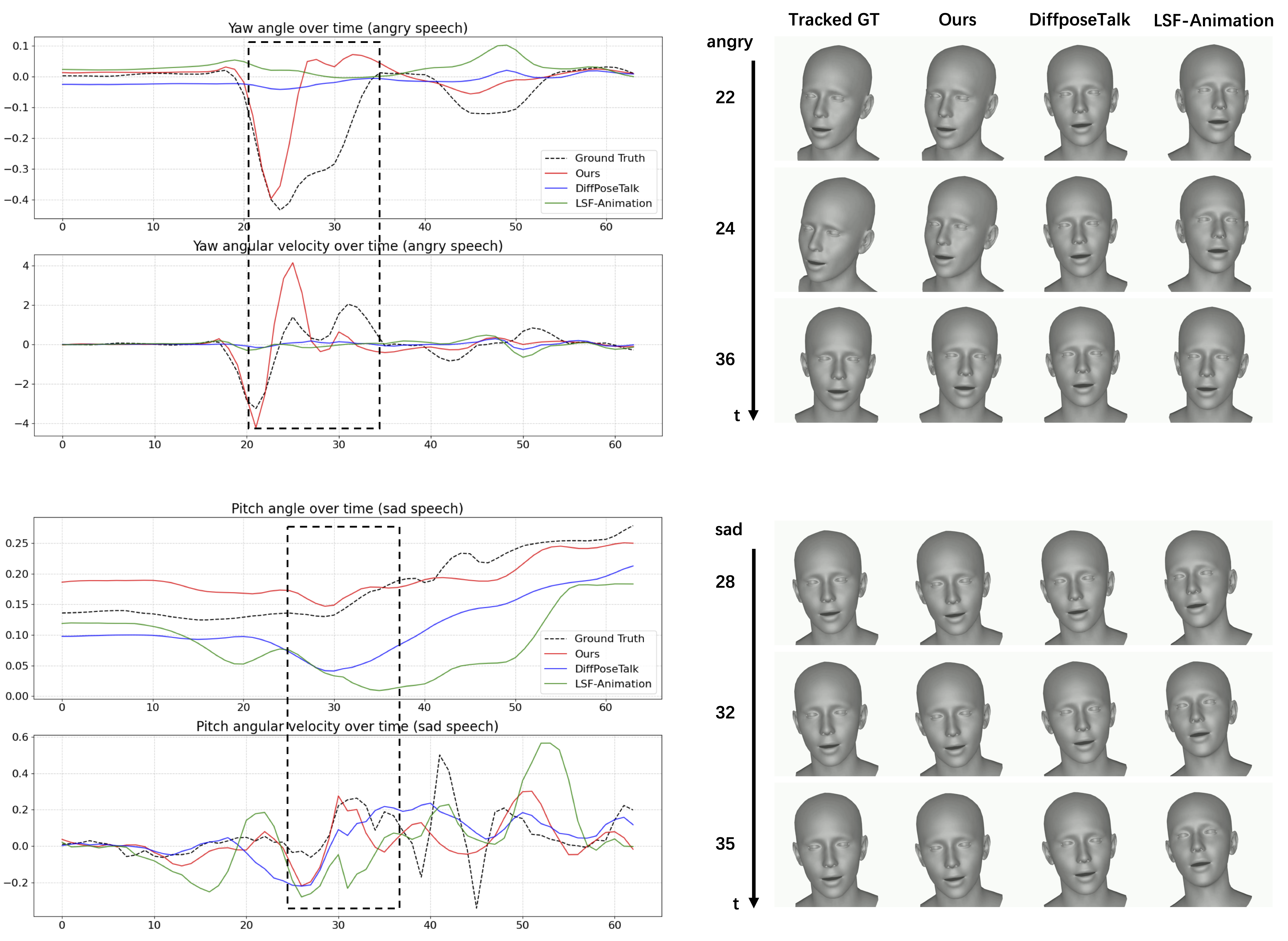}
    \caption{\textbf{Visualization of head pose dynamics.} The left line charts display the temporal evolution of pitch and yaw angles alongside their corresponding angular velocities. Notable segments are highlighted with dashed boxes, represented by the rendered head poses on the right. Compared with baselines, our method produces motion patterns that are more consistent with speaker-specific styles under emotional speech conditions.}
    \label{fig:head-pose}
\end{figure*}

\begin{table*}[tbp]
\caption{Quantitative comparison with state-of-the-art methods on the 3D-RAVDESS dataset.
The global best results are shown in \textbf{bold}. For each compared method, the better result between its original version and the variant augmented with our proposed motion-aligned speech encoder is \underline{underlined}.}
\label{tab:in domain}
\centering
\begin{threeparttable}
\begin{tabular}{lccccc}
\toprule
\multicolumn{1}{c}{\multirow{2}{*}{Method}} & \multicolumn{3}{c}{Facial Expression} & \multicolumn{2}{c}{Head Pose} \\
\cmidrule(lr){2-4} \cmidrule(lr){5-6}
 & LVE($\times10^{-5}$mm)$\downarrow$ & EVE($\times10^{-6}$mm)$\downarrow$ & FDD($\times10^{-7}$mm)$\downarrow$ & BA($\times10^{-1}$)$\uparrow$ & FID($\times10^{-2}$)$\downarrow$ \\
\midrule
{DiffPoseTalk}$^{\ast}$ & {2.554} & {1.276} & \textbf{12.770} & {1.924} & {5.780} \\
DiffPoseTalk & {3.686} & {2.223} & {13.980} & {1.487} & {13.320} \\
DiffPoseTalk+enh. & {\underline{3.136}} & {\underline{1.945}} & {\underline{13.900}} & {\underline{1.688}} & {\underline{7.150}} \\
\midrule
LSF-Animation  & {3.300} & 2.206 & 14.550 & 1.784 & 6.452 \\
LSF-Animation+enh. & \underline{3.258} & {\underline{1.712}} & {{\underline{14.052}}} & {\underline{1.880}} & {\underline{5.867}} \\
\midrule
Ours & \textbf{2.366} & \textbf{1.266} & 13.660 & \textbf{1.946} & \textbf{2.910} \\
\bottomrule
\end{tabular}

\begin{tablenotes}
      \footnotesize
        \item ${\ast}$ indicates the oracle version, which utilizes the actual talking style of the target speaker as an extra condition alongside speech.
\end{tablenotes}
\end{threeparttable}
\end{table*}
\par

\textbf{\textcolor{black}{Out-of-Domain Evaluation.}}
The relevant quantitative comparisons are reported in Table~\ref{tab:cross domain}. Our method consistently outperforms DiffPoseTalk and LSF-Animation across all metrics. Notably, it also surpasses DEEPTalk, despite the latter being evaluated in-domain, whereas our approach, together with DiffPoseTalk and LSF-Animation, operates under a more challenging \textcolor{black}{out-of-domain} setting. This result demonstrates the strong generalization capability of our framework. Comparing in-domain and \textcolor{black}{out-of-domain} performance, we observe a clear trend of performance degradation, most notably on LVE for lip synchronization. This is likely attributed to the limited scale and phonetic diversity of the 3D training data, which constrain the coverage of speech patterns in unseen domains.
\begin{table*}[tbp]
\centering
\caption{Quantitative comparison with state-of-the-art methods on the 3D-MEAD dataset. Best results are highlighted in \textbf{bold}. Head pose metrics for DEEPTalk are omitted since the method does not generate head motion.}
\begin{tabular}{lccccc}
\toprule
\multicolumn{1}{c}{\multirow{2}{*}{Method}} & \multicolumn{3}{c}{Facial Expression} & \multicolumn{2}{c}{Head Pose} \\
\cmidrule(lr){2-4} \cmidrule(lr){5-6}
 & LVE($\times10^{-5}$mm)$\downarrow$ & EVE($\times10^{-6}$mm)$\downarrow$ & FDD($\times10^{-7}$mm)$\downarrow$ & BA($\times10^{-1}$)$\uparrow$ & FID($\times10^{-2}$)$\downarrow$ \\
\midrule
DiffPoseTalk & {19.88} & {3.543} & {8.633} & {2.592} & {3.798} \\
LSF-Animation  & {19.10} & 3.505 & 8.456 & 2.578 & 7.989 \\
DEEPTalk & {19.98} & {3.320} & {10.505} & {-} & {-} \\
Ours & \textbf{17.59} & \textbf{3.263} & \textbf{8.298} & \textbf{2.601} & \textbf{3.200} \\
\bottomrule
\end{tabular}
\label{tab:cross domain}
\end{table*}
\par

\textbf{\textcolor{black}{Inference speed.}}
The inference efficiency is reported in Table \ref{tab:speed comparison}.  For a fair comparison, all methods are evaluated on the same hardware setup using a single NVIDIA RTX 3090 GPU. Our approach achieves an inference speed of 20.48 FPS. While slightly slower than DiffPoseTalk, it surpasses LSF-Animation and satisfies the requirements for near real-time applications.
\begin{table}[tbp]
\caption{Comparison of inference speed (FPS) among different methods.}
\label{tab:speed comparison}
\centering
\begin{tabular}{lccc}
\toprule
Method & DiffPoseTalk & LSF-Animation & Ours \\
\midrule
FPS $\uparrow$ & \textbf{27.39} & 19.36 & 20.48 \\
\bottomrule
\end{tabular}
\end{table}

\subsection{Qualitative Comparisons}
\begin{figure*}[tbp]
    \centering
    \includegraphics[width=1\textwidth]{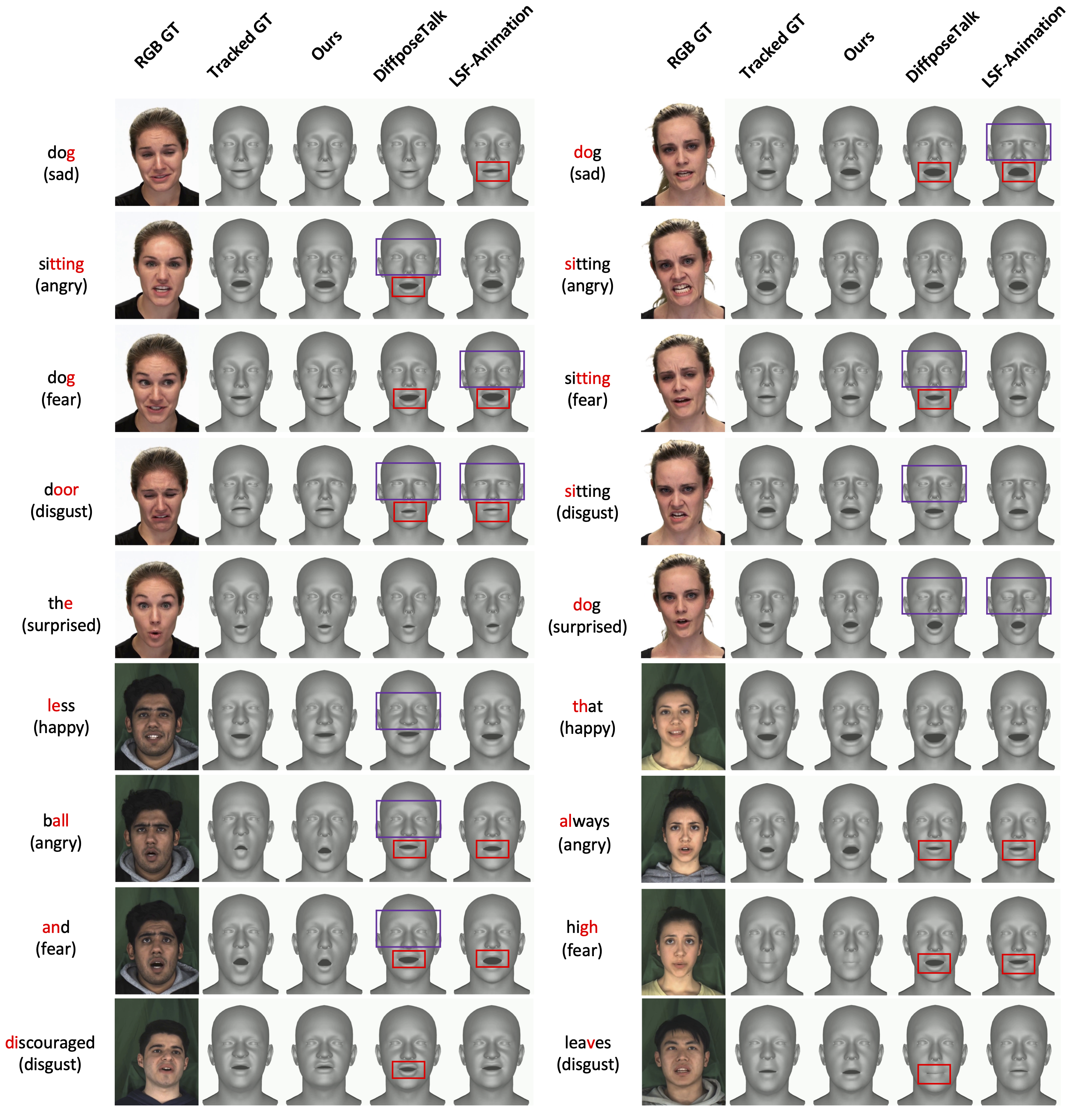}
    \caption{\textbf{Visual comparison of facial motion during emotional speech.} To facilitate comparison, all results are rendered in a canonical frontal view with zero head pose. The top five rows present in-domain results on 3D-RAVDESS, while the bottom four rows show out-of-domain results on 3D-MEAD. Purple boxes highlight upper-facial inconsistencies such as emotional conflicts or deviant styles, while red boxes indicate lip motion errors. Compared to baselines, our approach captures speaker-specific styles and emotional nuances more accurately, demonstrating superior generalization on unseen data.}
    \label{fig:visualization}
\end{figure*}

Figure~\ref{fig:visualization} compares facial motions generated by our method against competing approaches. In in-domain evaluations, while most methods capture coarse emotional content, our method uniquely preserves speaker-specific styles and fine-grained nuances, such as the subtle downward mouth curvature typical of disgust. In \textcolor{black}{out-of-domain} settings, performance naturally degrades across all models due to the domain gap. However, our approach demonstrates greater resilience. Competitors occasionally suffer from severe emotional inconsistency. For instance, DiffPoseTalk erroneously pairs happy expressions with fearful speech. Moreover, when articulation is jointly constrained by emotion and content, baselines like DiffPoseTalk and LSF-Animation show severe lip–speech asynchrony. In contrast, our method mitigates these artifacts, maintaining high fidelity and synchronization. Finally, as shown in Fig.~\ref{fig:head-pose}, our model generates head pose dynamics that remain coherent with the speaker’s idiosyncratic patterns across emotions. Please refer to the accompanying demo video for a comprehensive evaluation.
\par

\subsection{Analysis of Motion-Aligned Speech Encoder}
\textbf{Method-Agnostic Enhancement.}
Our motion-aligned speech encoder is designed as a standalone component that can be seamlessly integrated into existing pipelines after pre-training. To evaluate its generality, we incorporate the encoder into two representative baseline methods and report the results in Table~\ref{tab:in domain}. Across most evaluated metrics, the augmented variants consistently outperform their original counterparts, with particularly notable gains on metrics related to facial expressiveness and head motion dynamics. These improvements indicate that the proposed speech encoder contributes complementary motion-relevant information beyond what is captured by the original architectures, and can serve as a method-agnostic enhancement. While we focus on two baselines for clarity, extending this integration to a broader range of 3D talking head frameworks is a promising direction for future work.
\par

\begin{figure*}[t]
    \centering
    \subfloat[]{\includegraphics[width=0.35\textwidth]{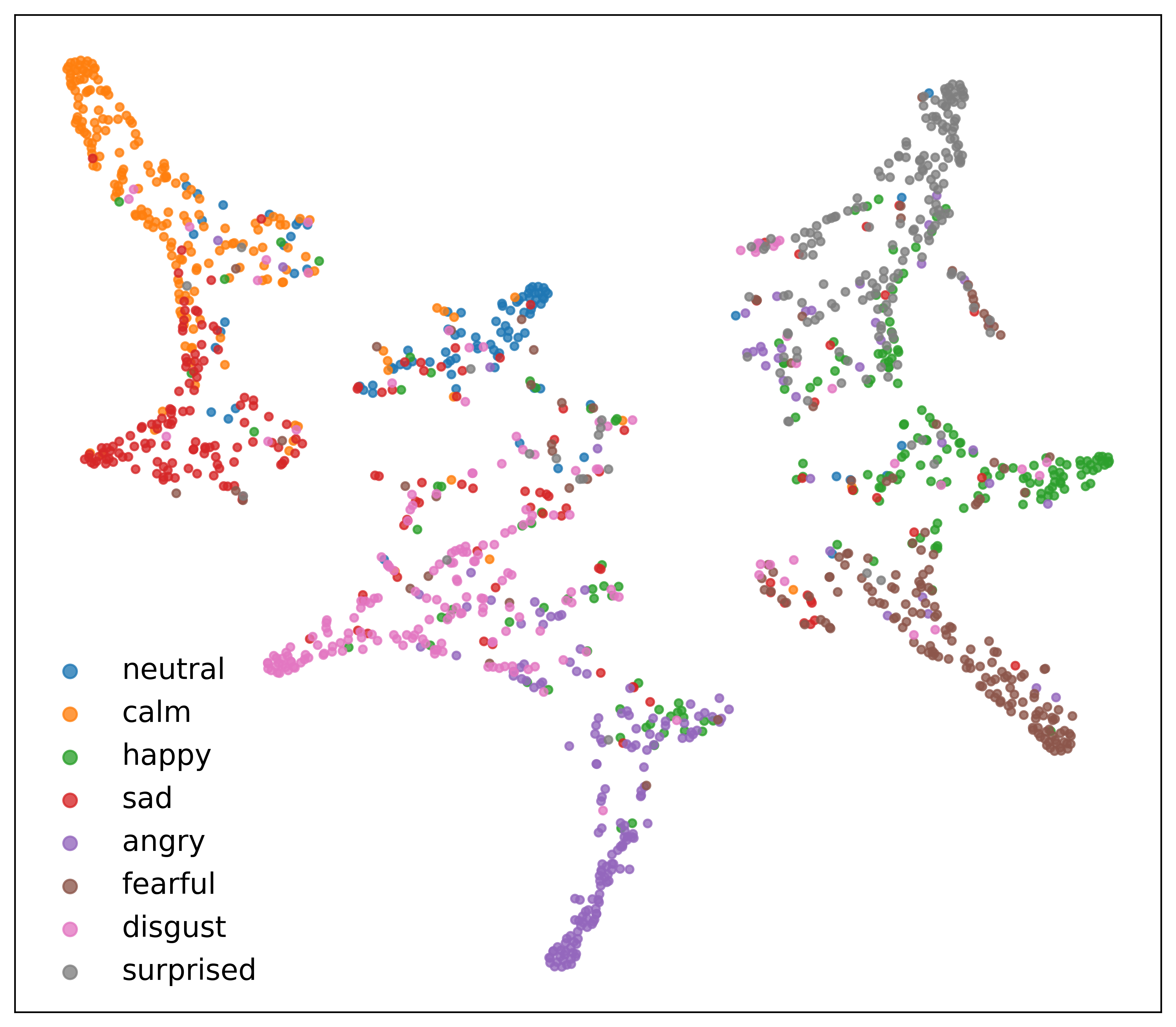}
        \label{fig:a}
    }
    \subfloat[]{\includegraphics[width=0.35\textwidth]{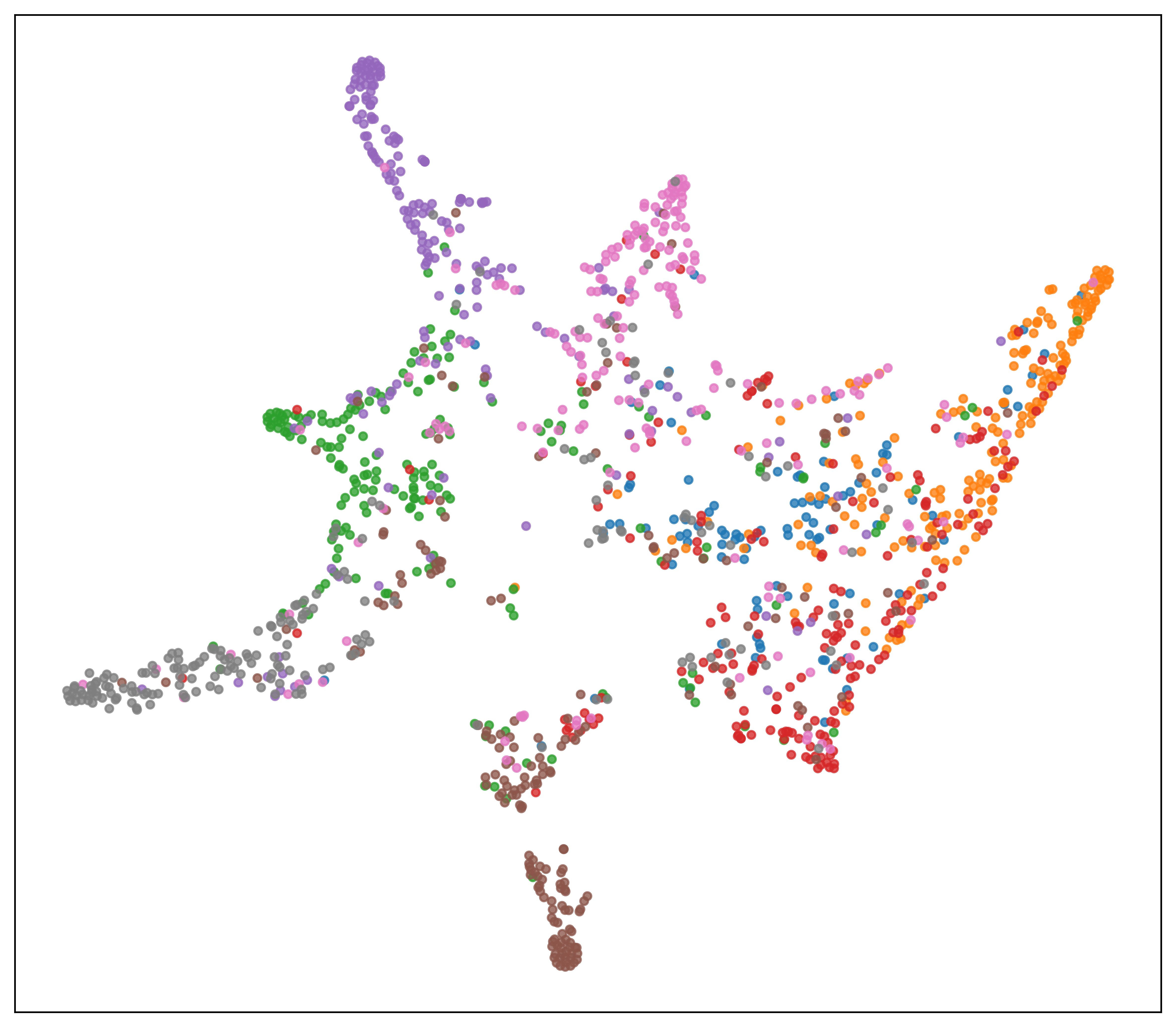}
        \label{fig:b}
    }
    \caption{t-SNE visualization of learned representations on the RAVDESS dataset. (a) Representations extracted by our motion-aligned speech encoder trained with audio-visual data. (b) Representations extracted by the audio-only variant.}
    \label{fig:tsne}
\end{figure*}

\textcolor{black}{\textbf{Visualizing the Learned Representation Space.}
To evaluate the motion-aligned speech encoder, we train a two-layer MLP classifier on its frozen representations using the RAVDESS dataset with an 8:2 train-test split. For comparison, we construct an audio-only variant of the proposed speech encoder, which shares the same configuration but is pre-trained without paired visual data.
The t-SNE visualizations of the extracted features reveal several key properties of the learned representation space. First, as shown in Fig. \ref{fig:tsne}(a), the features extracted by our motion-aligned speech encoder place psychologically similar emotions closely in the latent space. Specifically, low-arousal emotions (e.g., calm, sad, and disgusted) are primarily distributed on one side, whereas high-arousal emotions (e.g., happy, surprised, and fearful) cluster on the other. In contrast, this arousal-based separation is much less apparent in the audio-only variant, as shown in Fig. \ref{fig:tsne}(b). This structural difference highlights the enhanced discriminative capability introduced by paired visual supervision during pre-training.
Second, upon closer inspection of Fig. \ref{fig:tsne}(a), we further observe that samples from the same emotion category tend to form coherent clusters and continuous manifolds, indicating that the encoder effectively captures high-level expressive cues from speech. Furthermore, the smooth transitions and overlaps between different emotion clusters reflect the inherent continuity of human expression. Effectively modeling these blended states results in a robust continuous manifold, which is essential for generating natural 3D talking head animations and mitigating the risk of motion jitter.}


\par

\textcolor{black}{\textbf{Scalability Analysis.}}
\textcolor{black}{We further investigate the scalability of the proposed motion-aligned speech encoder by varying the volume of pre-training data. 
As shown in Table \ref{tab:scaleup}, compared with a baseline that relies solely on WavLM and emotion2vec, the introduction of our encoder yields substantial improvements across all metrics. However, when scaling up the pre-training data across three curated datasets of increasing sizes, we observe that subsequent performance gains rapidly plateau and become marginal.
The reasons behind these diminishing returns are twofold. First, the initial performance leap stems from bridging a critical modality gap. While WavLM and emotion2vec extract rich semantic and emotional features, they fundamentally lack awareness of facial and head motion dynamics. Our encoder successfully injects these missing audio-motion co-occurrence priors into the system.
Second, the limited gains from further scaling suggest that learning generic audio-motion alignment is highly sample-efficient. Unlike open-ended semantic learning, the mapping between speech and kinematic dynamics is relatively constrained. Once exposed to a moderate amount of paired data, the model adequately captures the essential alignment priors. Consequently, the headroom for further enhancing the alignment representation via data scaling may be smaller than initially hypothesized.
Ultimately, this observation raises an open question regarding scaling laws in speech representation learning: to achieve optimal downstream performance, is it more effective to scale multimodal pre-training (exploiting audio-motion priors) or to scale unimodal unsupervised pre-training (learning from massive raw audio)? We leave this investigation to future work.}
\par

\begin{table*}[tbp]
\centering
\caption{
Quantitative results under different pre-training data scales. Baseline denotes the basic 3D talking head animation framework introduced in Section 3.2 of the main text without the motion-aligned speech encoder. Three sets denote variants using motion-aligned speech encoders pre-trained on different scales of 2D audio-visual data. To account for training randomness, we report the mean and standard deviation across three independent runs for each variant. Best results are highlighted in \textbf{bold}.
}
\begin{tabular}{lccccc}
\toprule
\multicolumn{1}{c}{\multirow{2}{*}{Method}} & \multicolumn{3}{c}{Facial Expression} & \multicolumn{2}{c}{Head Pose} \\
\cmidrule(lr){2-4} \cmidrule(lr){5-6}
 & LVE($\times10^{-5}$mm)$\downarrow$ & EVE($\times10^{-6}$mm)$\downarrow$ & FDD($\times10^{-7}$mm)$\downarrow$ & BA($\times10^{-1}$)$\uparrow$ & FID($\times10^{-2}$)$\downarrow$ \\
\midrule
Baseline & {2.787} & {1.340} & {14.980} & {1.574} & {3.858} \\
Small set
 & {2.413$\pm$0.050} & {1.303$\pm$0.038} & {13.778$\pm$0.139} & {1.971$\pm$0.223} & {3.077$\pm$0.255} \\
Medium set
  & {2.371$\pm$0.013} & {1.243$\pm$0.012} & {13.337$\pm$0.107} & {2.027$\pm$0.065} & {2.898$\pm$0.012} \\
Large set & \textbf{2.357$\pm$0.021} & \textbf{1.205$\pm$0.025} & \textbf{11.922$\pm$1.026} & \textbf{2.168$\pm$0.096} & \textbf{2.837$\pm$0.022} \\
\bottomrule
\end{tabular}
\vspace{-5pt}
\label{tab:scaleup}
\end{table*}


\subsection{User Study}
To validate perceptual performance, we recruited twenty-six participants to rate our method against baselines across four complementary criteria. To prevent confounding factors, we employed specific rendering strategies for different metrics: (1) \textbf{Lip-Sync Quality} and \textbf{Expression Naturalness} were evaluated on zero-pose meshes to strictly isolate facial articulation; (2) \textbf{Head Motion Naturalness} was assessed on meshes \textcolor{black}{with all facial animation disabled to focus on global pose dynamics}; and (3) \textbf{Overall Naturalness} was rated on full animation sequences. \textcolor{black}{Driven by both in-domain and out-of-domain speech, these animations were presented to participants side-by-side in a randomized order (our method vs. a baseline). All evaluations were scored using a well-established five-point Likert scale \cite{likert1932technique}.}



\begin{figure}[tbp]
    \centering
    \includegraphics[width=\linewidth]{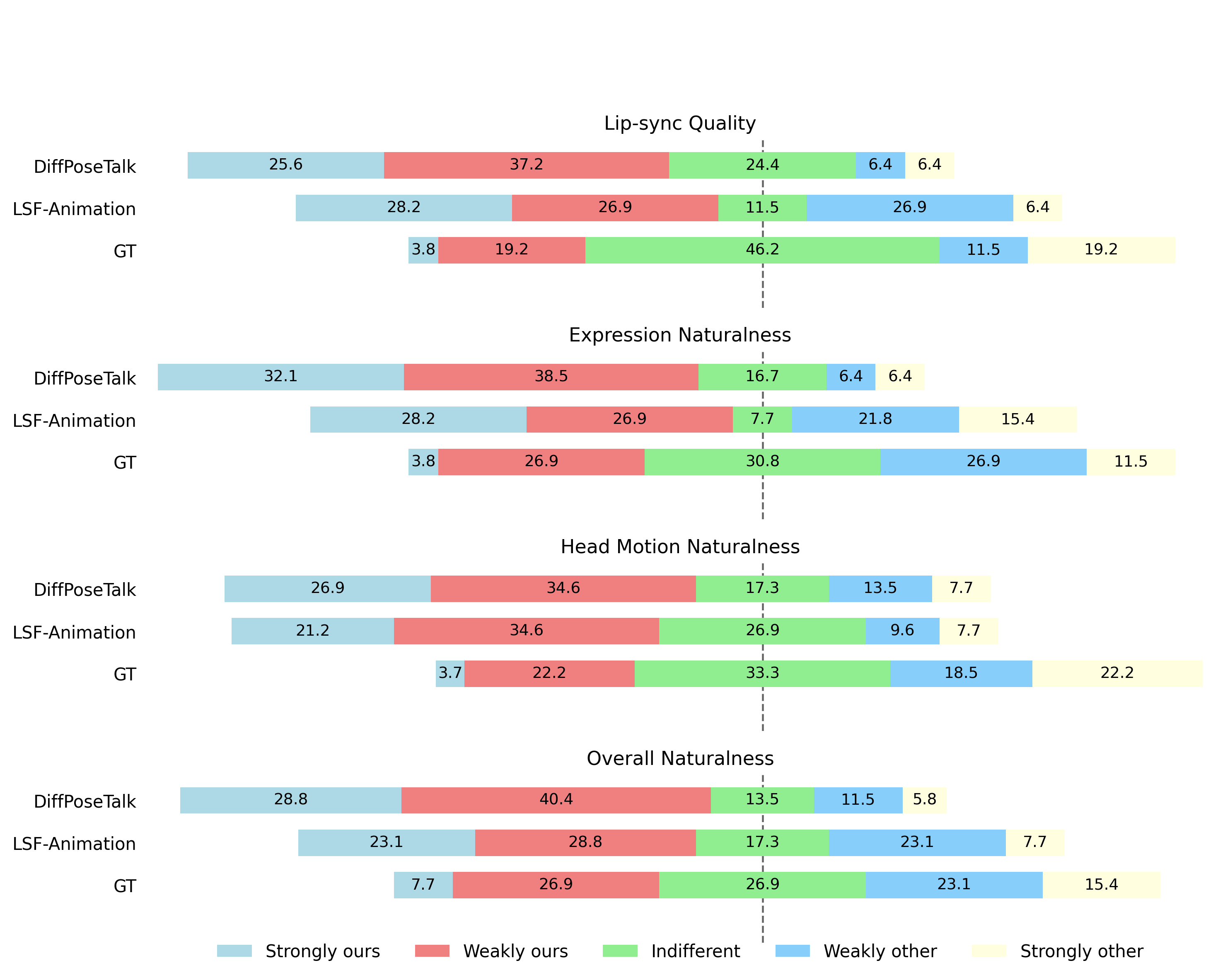} 
    \caption{\textbf{User study results.} Participants were asked to indicate their preferred result between our method and a competing approach under the specified evaluation criteria. The chart reports the corresponding preference percentages.}
    \label{fig:user-study}
\end{figure}

\textcolor{black}{
As shown in Fig. \ref{fig:user-study}, our method consistently outperforms baseline approaches across all four perceptual criteria, achieving over 65\% and 54\% user preferences against DiffPoseTalk and LSF-Animation, respectively. More importantly, when compared to the Ground Truth, approximately 60\% of participants rated our results as either indistinguishable from or even preferable to the tracked human motions in facial expressions, head movements, and overall dynamics. These findings confirm that our approach bridges the perceptual gap between synthesized animations and real-world human behavior.
}
\par

\subsection{Ablation Analysis}
\label{ablation_study}
\textcolor{black}{Due to space limitations, only quantitative ablation results are included here. Please refer to the supplementary video for visual comparisons.}
\begin{table*}[tbp]
\caption{Quantitative results of ablation studies on the 3D-RAVDESS dataset. The upper section reports variants of our self-distillation framework for learning the motion-aligned speech encoder, the middle section reports the results of combining this encoder with existing representations to demonstrate its complementary role, and the lower section reports variants in how the learned encoder is applied to the 3D talking head model to enhance expressiveness. Best results are highlighted in \textbf{bold}.}
\centering
\begin{tabular}{lccccc}
\toprule
\multicolumn{1}{c}{\multirow{2}{*}{Method}} & \multicolumn{3}{c}{Facial Expression} & \multicolumn{2}{c}{Head Pose} \\
\cmidrule(lr){2-4} \cmidrule(lr){5-6}
 & LVE($\times10^{-5}$mm)$\downarrow$ & EVE($\times10^{-6}$mm)$\downarrow$ & FDD($\times10^{-7}$mm)$\downarrow$ & BA($\times10^{-1}$)$\uparrow$ & FID($\times10^{-2}$)$\downarrow$ \\
\midrule
w/o vis. info.
 & {2.896} & {1.653} & {13.850} & {1.743} & {3.832} \\
w/o aud. mask.
  & {2.866} & {1.348} & {13.730} & {1.757} & {3.421} \\
w/o vis. mask. & {2.443} & {1.286} & {14.370} & {1.830} & {3.814} \\
w/o dropout & {2.747} & {1.649} & {13.730} & {1.797} & {3.790} \\
Ours full & \textbf{2.366} & \textbf{1.266} & \textbf{13.660} & \textbf{1.946} & \textbf{2.910} \\
\midrule
WavLM & {3.686} & {2.059} & {13.790} & {1.492} & {3.528} \\
WavLM+emotion2vec & {2.787} & {1.340} & {14.980} & {1.574} & {3.858} \\
WavLM+MASE & {3.115} & {1.601} & {13.860} & {1.820} & {3.088} \\
Ours full & \textbf{2.366} & \textbf{1.266} & \textbf{13.660} & \textbf{1.946} & \textbf{2.910} \\
\midrule
w/o in. enh. \& out. sup. & {2.787} & {1.340} & {14.980} & {1.574} & {3.858} \\
\textcolor{black}{F0 mod. emotion2vec} & {3.044} & {1.477} & {13.768} & {1.784} & {3.832} \\
w/o in. enh. & {2.557} & {1.317} & {14.450} & {1.704} & {3.812} \\
w/o out. sup. & {2.579} & {1.387} & {14.410} & {1.839} & {3.708} \\
Ours full & \textbf{2.366} & \textbf{1.266} & \textbf{13.660} & \textbf{1.946} & \textbf{2.910} \\
\bottomrule
\end{tabular}
\label{tab:ab1}
\end{table*}

\textbf{Motion-Aligned Speech Encoder Learning.} 
To validate the design choices of our motion-aligned speech encoder, we conduct an ablation study with four representative variants, applying modifications symmetrically to both student and teacher networks. The setup includes: (\romannumeral1) excluding the visual branch entirely; (\romannumeral2) \& (\romannumeral3) independently removing the emotion-aware masking mechanism in visual and audio branches; and (\romannumeral4) omitting the dropout operation during feature fusion. Quantitative results in the upper part of Table~\ref{tab:ab1} reveal three critical insights. First, relying solely on audio signals causes the most significant performance drop, confirming that visual cues provide indispensable motion and emotion priors that mitigate audio ambiguity. Second, disabling emotion-aware masking in either modality degrades performance, suggesting that these mechanisms effectively drive the network to learn robust, context-aware representations rather than overfitting to local patterns. Finally, the absence of fusion dropout harms downstream animation quality, verifying its role in preventing modality dominance and enhancing the encoder's generalization capability. 
\par


\textbf{Complementarity of the Motion-Aligned Speech Encoder (MASE).} 
MASE is designed to complement rather than replace existing speech representations by providing explicit motion-aligned cues. Since WavLM already provides a robust general-purpose speech representation, we establish it as our baseline and progressively integrate emotion2vec and MASE to investigate their synergistic effects. As presented in the middle section of Table~\ref{tab:ab1}, the addition of emotion2vec mainly enhances facial expressiveness (EVE), confirming that richer emotion representations benefit expression generation. However, it degrades facial dynamics (FDD) and head plausibility (FID), suggesting that emotion information alone is insufficient to determine when facial and head motions should occur. 
By contrast, adding MASE yields more consistent improvements, especially in head motion quality (BA and FID), while also improving facial metrics. This is consistent with our motivation that MASE learns motion-aligned speech cues for modeling speech-motion correspondence.
The integration of all three representations achieves state-of-the-art performance across all metrics, elegantly confirming their specific roles: WavLM captures \textit{what} is spoken, emotion2vec captures \textit{how} it is spoken, and MASE captures \textit{when} the facial and head movements should occur.

\textbf{Dual Guidance for Expressiveness Enhancement.}  
To further evaluate the efficacy of the proposed motion-aligned speech encoder, we conduct an ablation study with variants of the full model. Quantitative results are reported in the bottom section of Table~\ref{tab:ab1}. The first variant represents the baseline framework (Section \ref{talking head model}) without any guidance. 
\textcolor{black}{The second variant directly uses normalized F0 to modulate emotion2vec features. Compared to the baseline, it yields improvements in movement expressiveness and rhythmic alignment, but with a clear degradation in lip-sync performance. 
The third variant shares the same modulation mechanism, instead replacing the raw F0 signal with features extracted by the proposed speech encoder.
Contrasting their respective impacts relative to the baseline highlights the superiority of our learned speech features: whereas raw F0 forces a trade-off by sacrificing lip-sync for expressiveness, our encoder successfully improves both. 
Meanwhile, the fourth variant incorporates the speech encoder solely at the output supervision stage. Compared to the baseline, both the third and fourth single-guidance variants yield consistent performance improvements across all metrics. Notably, the full model, which combines input- and output-level guidance, achieves the most significant gains, demonstrating a clear complementary effect between the two.
}
\par


\section{Limitations and Future Work}
\textcolor{black}{Despite its effectiveness, our approach has several limitations that suggest directions for future work. First, our current masking strategy relies solely on F0, which may limit the discrimination of emotions that share similar pitch but differ in energy or spectral cues. Second, the expressiveness of our model is partially constrained by the pseudo-GT from monocular reconstruction, which often underestimates subtle facial motions. 
Third, while our method demonstrates improved generalization, it may struggle in extreme in-the-wild scenarios where the emotional tone of the speech is ambiguous or barely recognizable, leading to facial and head motions that are inconsistent with the intended emotion.
Furthermore, the computational overhead of the diffusion-based framework motivates the exploration of more efficient schemes, such as optimized autoregressive modeling \cite{DBLP:conf/nips/LiTLDH24} or flow matching \cite{DBLP:conf/iclr/LipmanCBNL23}. Finally, extending the framework to better capture long-term emotional transitions remains a key challenge to be addressed.}

\section{Conclusion}

We present ETHead to synthesize expressive 3D facial and head dynamics from speech, mitigating the scarcity of high-fidelity 3D data.
Our method employs a self-distillation framework with emotion-modulated masking, training a motion-aligned speech encoder on large-scale 2D videos. 
Experiments confirm that applying this specialized encoder with dual guidance remarkably enhances facial vividness and natural head motions. 
Moreover, this encoder serves as a 
transferable primitive that can be readily integrated into existing pipelines to boost expressiveness.
\par


\bibliographystyle{IEEEtran}
\bibliography{references}


{\appendix
\section*{Speech Processing}
In practice, the F0 processing is conducted online. To isolate valid acoustic data, we first extract non-silent voiced frames (F0$>$0) and apply a logarithmic transformation. Since establishing a true emotional baseline (i.e., the mean and standard deviation of log-F0 in a neutral state) is challenging in in-the-wild scenarios, we approximate these statistics over long-term speech utterances. The log-transformed F0 values are then standardized using a z-score normalization based on these approximated baseline parameters. Ultimately, we compute the absolute values of the normalized F0 to quantify the magnitude of pitch deviation. This absolute deviation serves as a theoretical proxy for the divergence from the speaker's neutral state, yielding an explicit emotion saliency profile over time.

\section*{Emotion-Aware Audio Masking Mechanism}
Guided by the emotion saliency profile, we design a non-uniform masking strategy for the audio branch. In essence, it prioritizes emotionally salient segments while maintaining diversity through uniform sampling. Specifically, the input Mel-spectrogram is first tokenized into a grid of patches. To align the high-resolution saliency profile with this patch-based representation, a base salience score for each patch is derived by averaging the absolute standardized F0 values over the time frames it spans. Since prosody spans across frequencies, we broadcast the temporal saliency score along the frequency dimension. This ensures that all frequency bins within a high-saliency time step share the same high masking probability, preserving vertical spectral continuity. To balance the training signal, the final masking distribution is a mixture of the softmax-normalized saliency scores and a uniform distribution. By sampling from this hybrid distribution, the student network is \textcolor{black}{exposed to only partial prosodic cues} in emotionally salient segments, forcing it to infer these dynamics from the remaining acoustic context. 
\par

\section*{Emotion-Aware Visual Masking mechanism}
To enforce cross-modal consistency, we extend the prosody-driven masking logic to the visual domain. While the temporal saliency profile indicates \emph{when} to mask, the visual branch must further identify \emph{where} the prosodic dynamics are visually manifested. First, we encode the pre-computed absolute standardized F0 trajectory into a sequence of prosody tokens using a lightweight 1D convolutional encoder. We then compute the cross-modal correlation between each visual patch token and this global sequence of prosody tokens. A high correlation score implies that a visual region is strongly aligned with the overall prosodic dynamics of the clip. Leveraging these alignment scores, we employ a ranking-based strategy: visual patches are sorted in descending order of correlation, and the top-ranked regions are selected as primary candidates for masking. Similar to the audio branch, we introduce stochasticity by randomly preserving a fraction of salient patches, preventing the model from overfitting to fixed correlation patterns.

\section*{Metrics}
To quantitatively evaluate the quality of synthesized 3D facial motion, we adopt three standard metrics following previous conventions: Lip Vertex Error (LVE) \cite{DBLP:conf/iccv/RichardZWTS21}, Emotional Vertex Error (EVE) \cite{DBLP:conf/iccv/PengWSXZH0F23}, and Upper-Face Dynamics Deviation (FDD) \cite{DBLP:conf/cvpr/XingXZC0W23}. Note that all three metrics are computed after transforming vertices to a zero head pose. LVE measures the maximum L2 error of lip vertices between the predicted and ground-truth meshes for each frame, reported as the average over all test frames. Similarly, EVE is defined as the average maximum L2 error over vertices in the eye and forehead regions. FDD evaluates upper-face dynamics by computing the average absolute difference between the temporal standard deviations of vertex positions in the generated and ground-truth animations. Among these, LVE primarily reflects lip synchronization, while EVE and FDD are indicative of emotional congruency. For head pose evaluation, we employ  Beat Alignment (BA) \cite{sun2024diffposetalk} and Frechet Inception Distance (FID) \cite{DBLP:conf/cvpr/SiyaoYGLW0L022}. BA quantifies rhythmic consistency by calculating the average temporal distance between each synthesized head-movement beat and its nearest counterpart in the ground-truth sequence. To assess the realism of the generated head poses, we compute the FID between the distributions of synthesized and real sequences based on kinetic features \cite{DBLP:conf/eurographics/OnumaFH08}.

\section*{Implementation Details}
Our method is implemented in PyTorch, with all experiments conducted on a single NVIDIA RTX 3090 GPU. In the pre-training phase, the backbones for both the audio and visual branches comprise a stack of 12 Transformer blocks, each featuring 12 attention heads. We employ the AdamW optimizer with a learning rate of $1\times10^ {-4}$ and a batch size of 64. The masking ratio for the audio branch is fixed at 0.6, whereas the visual branch utilizes a dynamic masking strategy, gradually increasing from 0.1 to 0.6. 
\textcolor{black}{The momentum $\alpha$ for updating the teacher network is set to 0.996, and the dropout rate $\beta$ for visual features to 0.3.}
To construct the joint speech-motion latent space, we freeze the pre-trained speech encoder and train the motion encoders, both consisting of 2 Transformer blocks with 12 attention heads. This training phase takes approximately 3 hours with a batch size of 64. For the 3D talking head model, both the facial motion and head pose generators employ a stack of 8 Transformer blocks with 8 heads serving as the diffusion denoising network. We process the input speech using a sliding window with $T_w=32$ and $T_p=4$. The diffusion process adopts a cosine noise schedule with 500 diffusion steps. These denoising networks are trained for 50k iterations with 5k warmup steps and a batch size of 64. 
\textcolor{black}{During training, the CNN feature extractor of WavLM is frozen, while its Transformer blocks are fine-tuned. For emotion2vec, only the last 6 Transformer blocks are fine-tuned, with the remaining components frozen.} 
The total loss function incorporates vertex reconstruction and geometric motion constraints. We empirically set the weights to $\lambda_{FM}^{vtx}=2\times10^6$, $\lambda_{FM}^{vel}=1\times10^7$, $\lambda_{FM}^{acc}=1\times10^5$, $\lambda_{HM}^{pos}=0.5$, $\lambda_{HM}^{vel}=50$, $\lambda_{HM}^{acc}=5$. respectively. This training stage requires approximately 16 hours.
}

\vfill

\end{document}